\documentclass[11pt,letterpaper]{article}

\usepackage[margin=1in]{geometry}
\usepackage{hyperref}
\usepackage{hyperref,enumitem}
\hypersetup{
	colorlinks=true,
	citecolor = blue       %
}
\setlist{noitemsep,leftmargin=\parindent,topsep=2pt}
\setlist{noitemsep,topsep=2pt}
\usepackage{natbib}
\usepackage{url}            %
\usepackage{xspace}
\usepackage{graphicx}
\usepackage{amsmath,amssymb,amsthm}
\usepackage{thm-restate}
\usepackage{booktabs} %
\usepackage{color}
\usepackage[font=small,labelfont=bf]{caption}
\usepackage{subcaption}
\usepackage{multirow}
\usepackage{authblk}

\usepackage{comment}

\usepackage[ruled]{algorithm2e} %

\SetAlFnt{\small}
\SetAlCapFnt{\small}
\SetAlCapNameFnt{\small}
\SetAlCapHSkip{0pt}
\IncMargin{-\parindent}

\DeclareMathOperator*{\argmin}{arg\,min}
\DeclareMathOperator*{\argmax}{arg\,max}

\newcount\Comments 
\newcommand{\E}{\mathbb{E}}
\newcommand{\1}{\mathbb{I}}
\newcommand{\I}{\mathbf{I}}
\newcommand{\eps}{\varepsilon}

\newcommand{\R}{\mathbb{R}}
\newcommand{\B}{\mathcal{B}}

\newcommand{\rev}{\mathbf{Rev}}

\newcommand{\PP}{\mathbb{P}}

\theoremstyle{plain}
\newtheorem{theorem}{Theorem}[section]
\newtheorem{corollary}{Corollary}[section]
\newtheorem{lemma}[theorem]{Lemma}
\newtheorem{definition}[theorem]{Definition}
\newtheorem{proposition}[theorem]{Proposition}

\newcommand{\new}[1]{{\color{black} #1}}

\begin{document}

\title{Robust Clearing Price Mechanisms for Reserve Price Optimization}

\author{Zhe Feng\thanks{Part of the work was done when the first author was a PhD student at Harvard University.}
}
\author{S\'ebastien Lahaie}

\affil{Google Research \authorcr \texttt{zhef@google.com, slahaie@google.com}}

\date{July 9, 2021}

\maketitle

\begin{abstract}
Setting an effective reserve price for strategic bidders in repeated auctions is a central question in online advertising. In this paper, we investigate how to set an anonymous reserve price in repeated auctions based on historical bids in a way that balances revenue and incentives to misreport. 
We propose two simple and computationally efficient methods to set reserve prices based on the notion of a clearing price and make them robust to bidder misreports. The first approach adds random noise to the reserve price, drawing on techniques from differential privacy. The second method applies a smoothing technique by adding noise to the training bids used to compute the reserve price.
We provide theoretical guarantees on the trade-offs between the revenue performance and bid-shading incentives of these two mechanisms. Finally, we empirically evaluate our mechanisms on synthetic data to validate our theoretical findings.
\end{abstract}

\section{Introduction}

A fundamental problem in auction design is the question of setting a reserve price to optimize revenue. In a data-rich environment like online advertising, it becomes possible to learn an effective reserve pricing policy by drawing on past bidding data~\citep{paes2016field}. %
Under simplifying assumptions such as i.i.d.\ bidders and a monotone hazard rate, the form of the optimal reserve price in a single-item auction has been well understood since the seminal work of~\citep{myerson1981optimal}. Reserve prices are still a revenue-optimal mechanism when these assumptions are relaxed, but they become challenging to implement both in theory~\citep{morgenstern2015pseudo} and in practice due to the nonconvexity of the optimization problem~\citep{medina2014learning}.

Nonetheless, in practice simple pricing strategies such as using some fixed quantile of historical bid distributions have been found to work well~\citep{ostrovsky2011reserve}, even if they are not technically optimal. In this paper, we consider a generalization of quantile-based reserve pricing recently proposed by~\citep{Shen19}, whereby a \emph{market-clearing} price fit to historical bid data is used as a reserve price going forward. Much previous work uses \emph{personalized} reserve prices, where each bidder sees a different price; this is largely motivated by the fact that when bidders values are not i.i.d.\ the optimal auction uses personalized reserves~\citep{myerson1981optimal}. The current standard in online display advertising, however, is to use an \emph{anonymous} reserve price common to all bidders. \citet{Shen19} show %
that fitting a clearing price to historical bids, using an appropriate convex loss function, leads to an effective anonymous reserve price with a favorable trade-off between revenue and match rate, compared to nonconvex surrogate losses that aim to directly optimize revenue~\citep{medina2014learning}.

However, a key concern with this approach (and with learning approaches more generally), is that bidders may be motivated to misreport if this leads to a more favorable reserve price in the future. That is, an auction such as a second price auction with fixed reserve, which is truthful in a single shot, may no longer be truthful in a repeated setting with dynamically updated reserve prices. Advanced bidders may take advantage of this fact~\citep{nedelec2019learning}. In this work, we therefore ask to what extent a reserve pricing policy such as clearing prices can be made \emph{robust} to bidder manipulations: bidders may still be able to effect future reserves, but the effect may be muted enough to make misreports hardly worthwhile. To add a measure of robustness to the mechanism, we consider two simple schemes, one based on techniques used in differential privacy~\citep{mcsherry2007mechanism} and the other based on smoothing techniques.
\textbf{Our Contributions.}
We propose two \emph{robust clearing price} (RCP) mechanisms: a differentially private RCP mechanism (DC-RCP) and a smoothing RCP mechanism (sRCP). We provide analytical characterizations of their performance to confirm that they can both effectively balance revenue and incentives.

For the DP-RCP mechanism, we first give a revenue guarantee in Theorem~\ref{thm:revenue-multiple-bidders} relating the performance of the noisy reserve to the original clearing price. %
To obtain a nuanced characterization of the incentives of DP-RCP, we consider an incentive compatibility metric recently proposed by~\citep{DLMZ20}. This metric is tailored to quantifying the incentives towards \emph{uniform bid shading}, a common strategy in online advertising where bids are simply values scaled down by a fixed constant factor. In Theorem~\ref{thm:ic-metric-multiple-bidders} we provide an \emph{exact} characterization of the IC metric of DP-RCP and find that it is directly related to the quantile chosen for pricing in the single bidder setting. 

DP-RCP requires random drawn noise for each auction, which may cause it to be problematic to use in practice. To handle this issue, we propose a smoothing RCP (sRCP) mechanism, where we add noise to the training bids to compute the reserve price. We characterize the revenue guarantee (Theorem~\ref{thm:smooth-revenue}) and IC-metric (Theorem~\ref{thm:smooth-ic-metric}) for the sRCP mechanism, in terms of the magnitude of the noise. \new{Theoretically, we show sRCP achieves better revenue guarantee ($O(1/\sqrt{\eps})$ revenue loss) than DP-RCP ($O(1/\eps)$ revenue loss), in terms of the magnitude of the noise $\eps$ in Theorem~\ref{thm:smooth-revenue}}. %

We validate our theoretical findings by simulating the two RCP mechanisms over synthetic bid data generated from lognormal distributions. We consider Laplace
noise to make the mechanisms robust.
\new{We find that the two RCP mechanisms have different tradeoff between revenue and IC-metric for different settings, e.g., whether to choose a conservative quantile to compute clearing price in the single-bidder case. For a suitably chosen parameter (i.e., $\lambda$, defined in Eq.~(\ref{eq:clean-clearing-price-loss}), which is used to control how aggressive the clearing price is, by generalizing the choice of quantile) %
for computing clearing price, sRCP achieves a better tradeoff between expected revenue and IC-metric than DP-RCP.}

\textbf{Related Work.} 
There is a rich and growing literature on learning algorithms for pricing in auctions. With respect to this paper, related work can be categorized along two dimensions: batch vs.\ online access to data, and whether bidders are myopic in their bidding. Our work considers batch learning of anonymous reserve prices under non-myopic bidders.

In the batch setting, starting with the work of~\citep{cole2014sample}, several works have studied the theoretical design of approximately optimal mechanisms based on historical samples of bidder values~\citep{balcan2019estimating,devanur2016sample,morgenstern2015pseudo}, with a focus on the complexity of the problem. A parallel line of research in machine learning develops learning algorithms for reserve prices based on historical batch samples of bids or values. \citep{medina2014learning,paes2016field} show how to learn reserve prices in lazy second price auctions, while~\citep{dutting19a} learn optimal multi-item auctions using deep learning, subject to incentive compatibility constraints. Most closely related to our work, \citep{Shen19} propose a reserve pricing policy based on learning the clearing price from historical bids. 
In~\citep{DengEtAl21} the authors examine the tradeoffs between revenue and incentives for different reserve pricing policies, just as we do in this work; whereas we consider smoothing techniques on a specific reserve pricing policy, their approach is to use mixtures of policies.
A loosely related work by \citep{maille2007} analyzed the equilibrium of clearing price in second price auctions.

All of these works assume that bidders are \emph{myopic}: they can strategize within individual auctions, but do not consider the impact of their bids on future auctions. Robustness to \emph{non-myopic} has recently been studied in~\citep{kanoria2014dynamic,epasto2018incentive}.
A common idea is to limit the influence of any particular bidder either by assuming a large number of bidders, or imposing a cost associated with manipulation. %
There has also been a lot of interest in the online version of the pricing problem~\citep{Caillaud04,amin2013learning,amin2014repeated,maocontextual}, but those ideas are not directly applicable to the batch-learning
setting. Robust learning is also relevant in the online model and
several papers develop policies that limit a non-myopic agent's incentive to
misreport~\citep{golrezaei2019dynamic,deng2019robustNon,liu2018learning,drutsa2018weakly,drutsa2020reserve}.

Our work applies ideas from a recent literature on metrics to quantify incentive compatibility. The most standard metric in this respect has been a bidder's regret from truthful bidding (i.e., the maximum foregone utility), which has been used as a design constraint in auctions where exact incentive compatibility was not possible or desirable~\citep{parkes2001achieving}. Computing regret as an informative metric in its own right is studied in~\citep{feng2019online,colini2020envy}. Due to its analytical tractability, we pay particular attention to the incentive metric introduced by~\citep{DLMZ20} when analyzing the robustness of our mechanism.
\section{Preliminaries}\label{sec:prelim}

We consider a setting with a set of $n$ bidders, denoted $[n] = \{1, 2, \dots, n\}$, participating in repeated auctions with a single seller.  The repeated auctions sell a single item at each round. The items arrive in an online manner and they must be sold once they arrive. For ease of exposition, we assume there is exactly one item per round.

At each round $t$,  each bidder $i$'s private valuation $v_{i,t} \in  V \subseteq\R_{\geq 0}$ is drawn independently from a distribution $D_{i, t}$ over $V$. Upon the arrival of the $t$-th item, each bidder $i$ observes her private valuation $v_{i, t}$ and submits a bid $b_{i, t}\in V$ to the seller. After receiving the bids $(b_{1, t}, b_{2, t}, \dots, b_{n, t})$ from all bidders at each round $t$, the seller runs an auction to decide how to allocate and charge for the item. Throughout this paper, we assume that $V$ is a closed interval, and without loss of generality rescale it to $V=[0, 1]$.

In this paper, we consider a widely adopted auction format: the \emph{second price auction with anonymous reserve price}. The seller allocates the item to the unique bidder who bids higher than the highest bid from all the other bidders $\max_{j\neq i} b_{j, t}$, 
and the reserve price $r_t$ at round $t$. (We assume that the bids are all distinct, which holds generically.) This winning bidder is charged $\max\{\max_{j\neq i} b_{j, t}, r_t\}$. %
In this paper, we focus on setting an anonymous reserve price, i.e., the reserve prices for all bidders are the same.

\subsection{Multi-stage Model for Reserve Pricing}
In this paper we consider a multi-stage model for strategic bidders. It follows a canonical setup: the seller learns an anonymous reserve price from the bidders' historical bids and applies it in the future. More specifically:
\begin{itemize}
\item In stage 1, the seller sets the reserve to 0.
\item For each stage $t > 1$, the seller computes a reserve $r_t$ from the bidders' bids in stage $t-1$. The seller sets the reserve $r_t$ to all the auctions in stage $t$.
\end{itemize}
We assume that each stage consists of sufficiently many auctions (queries) to faithfully estimate bid distributions. Throughout this paper, we further assume that for each bidder $i$, $D_{i, t} = D_i$ for all $t$. This \emph{stationarity} assumption of valuation widely holds for large markets in practice.

\subsection{Incentive Compatibility Metric}

In this paper, we adopt the dynamic incentive compatibility (dynamic-IC) metric proposed in~\citep{DLMZ20} to measure the incentive compatibility of the mechanism in the multi-stage model. A mechanism is dynamic-IC if for any stage $t$, reporting the true value at stage $t$ is an optimal strategy for the bidder regardless of her own value and others' bids, given the bidder always bid truthfully in the future \citep{DLMZ20}.

It is without loss of generality to assume there are only two stages as shown in~\citep{DengEtAl21}.\footnote{To elaborate on this, in our multi-stage model, when measuring
dynamic-IC for stage $t$, only stage $t+1$ is relevant since the
bids at stage $t$ only affect the reserve price at stage $t+1$; and
the reserve prices for all future stages are the same provided
that the buyer always reports truthfully from stage $t + 1$
onwards, under the dynamic-IC definition. Therefore, it is without loss of generality to focus on two-stage model.} It is well-known that the second price auction is incentive compatible (IC); i.e., reporting truthfully (bid is equal to value) is always a bidder's optimal bidding strategy, regardless of the others' bids and her private value. However, this no longer the case in the two-stage model, because the bidder may benefit from misreporting in stage 1 to induce the seller to set a lower reserve price in stage 2. In stage 2, given the reserve price, each bidder will report truthfully because of the truthfulness of the second price auction with fixed reserve.

In the two-stage model, we assume each bidder $i$ has a private (unknown to others and the seller) \emph{weakly increasing} bidding strategy $\beta_i: V \rightarrow V$, which maps values to bids in stage 1.
We denote by $D =\times_i D_i$ 
and $D_{-i} = \times_{j\neq i} D_j$
the distribution of valuation profile $v$ and the valuation profile $v_{-i}$ of the bidders other than $i$, respectively. Specifically, the bidding strategy $\beta_i$ is the identity function $\I_i$ if bidder $i$ reports truthfully. With a slight abuse of notation we write $\beta(v) = (\beta_1(v_1), \beta_2(v_2),\dots, \beta_n(v_n))$ and $\I(v) = (\I_1(v_1), \dots, \I_n(v_n))$. Let $\I_{-i}$ be the identity bidding strategies of the bidders other than $i$. Denote $v^{(1)}$ and $v^{(2)}$ be the largest and the second largest value among valuation profile $v$. Let $D^{(1)}$ and $D^{(2)}$ be the cumulative distribution function (CDF) of $v^{(1)}$ and $v^{(2)}$, respectively. Let $m_i = \max_{j\neq i} v_j$, as well as $g_i(\cdot)$ and $G_i(\cdot)$ are the corresponding probability density function (PDF) and CDF.

Formally, the reserve price set in stage 2 is a function of the distribution of the bids which are induced by the bidding function, so we write the reserve as $r(\beta)$.
We next define the expected utility for each bidder $i$ in both stages. In stage 1, %
let $\hat{u}_{i, 1}(\beta_i; v_i)$ be the expected utility of bidder $i$ in stage 1 when she adopts $\beta_i$ bidding strategy and the other bidders report truthfully, i.e.
\begin{eqnarray}\label{eq:expected-utility-stage-1}
\hat{u}_{i, 1}(\beta_i; v_i) = \E_{m_i\sim G_i}\left[(v_i - m_i)\cdot \1\{\beta_i(v_i) \geq m_i\}\right]
\end{eqnarray}
In stage 2, the optimal bidding strategy is to bid truthfully, given the reserve price computed from the bids in stage 1. Therefore, 
we define $\hat{u}_{i, 2}(\beta_i; v_i)$ as the expected utility of bidder $i$ in stage 2, when her bidding strategy is $\beta_i$ and the other bidders report truthfully in stage 1:
\begin{eqnarray}\label{eq:expected-utility-stage-2}
\hat{u}_{i, 2}(\beta_i; v_i) = \E_{m_i\sim G_i}\left(v_i - \max(m_i, r(\beta_i, \I_{-i}))\cdot \1\{v_i \geq \max(m_i, r(\beta_i, \I_{-i}))\right]
\end{eqnarray}
Given the above notations, we define the incentive compatibility metric (IC-metric) in the following,

\begin{definition}[IC-metric]\label{def:ic-metric}
The IC-metric for bidder $i$ locally at true value $v_i$ is
\begin{eqnarray*}
\mathtt{DIC}_i = 1 + \lim_{\alpha\rightarrow 0}\frac{\E_{v_i}\left[\hat{u}_{i, 2}(1+\alpha; v_i) - \hat{u}_{i, 2}(1-\alpha; v_i)\right]}{2\alpha\cdot E_{v}[v_i x_{i, 1}(v)]},
\end{eqnarray*}
where
$\hat{u}_{i, 2}(1+\alpha; v_i)$ represents the expected utility of bidder $i$ in stage 2 when her bidding strategy is $\beta_i(v_i) = (1+\alpha)v_i$ (resp. $\hat{u}_{i, 2}(1-\alpha; v_i)$), and $x_{i, 1}(v)$ is allocation of bidder $i$ under truthful bidding in stage 1. 
\end{definition}
To understand the intuition behind this metric, note that the second term in $\mathtt{DIC}_i$ is essentially the derivative of second-stage utility with respect to a uniform bid shading factor $\alpha$; i.e., it captures the marginal future utility gain from shading away from the true value $v_i$. Typically, bid shading is beneficial (because it lowers future reserves), in which case the second term in $\mathtt{DIC}_i$ is negative and the metric drops below 1. If truthful bidding (i.e., no shading) is optimal, then the second term is 0 and the metric takes on the reference value 1. Comparing to Regret, the IC-metric turns out to be much more analytically tractable because it is a local metric that focuses specifically on bid-shading deviations.

\subsection{Reserve Pricing via Clearing Prices}
There are many different approaches to setting reserve prices in second price auctions, most commonly focusing on \emph{personalized} reserves because these are revenue-optimal when bidders are not identical~\citep{myerson1981optimal}. A simple strategy for reserves considered in the literature is to use quantiles of historical bid distributions (e.g., setting the reserve price to the 25th quantile)~\citep{ostrovsky2011reserve}. In this paper, we focus on a reserve pricing strategy recently introduced by~\citet{Shen19} which generalizes using bid quantiles for individual bidders to an anonymous reserve price for all bidders. They propose to compute and apply the \emph{clearing price} of historical bid profiles as an anonymous reserve, obtained by optimizing the \emph{clearing loss} function
\begin{eqnarray}\label{eq:clean-clearing-price-loss}
\ell^c(p, b; \lambda) = \sum_{i=1}^n \max\{b_i - p, 0\} + \lambda p,
\end{eqnarray}
averaged over all historical bid profiles, where $b = (b_1,\dots,b_n)$.
The parameter $0\leq \lambda\leq n$ here controls the trade-off between match rate (probability that the item is sold) and revenue.

We consider using the clearing price to set the reserve price in stage 2 based on the bids from bidders in stage 1. Given the bidding strategies $\beta$ of bidders, the optimal clearing price is obtained as
\begin{eqnarray}\label{eq:optimal-expected-clearing-price}
p^c_\lambda(\beta) = \argmin_{p} \E_{v}\left[\ell^c(p, \beta(v); \lambda)\right].
\end{eqnarray}
The clearing loss is convex and can be easily optimized using standard machine learning libraries, e.g. stochastic gradient descent. Given a reserve price (clearing price) $r$ obtained by minimizing the clearing loss, we denote by $\rev(r)$ the expected revenue in stage 2 when bidders bid truthfully.

\begin{definition}[Expected Revenue]\label{def:revenue}
The expected revenue in stage 2 is defined as,
\begin{eqnarray*}
\rev(r) = \E_v\left[\max\{v^{(2)}, r\} \cdot \1\{v^{(1)} \geq r\}\right],
\end{eqnarray*}
where $v^{(1)}$ and $v^{(2)}$ are the largest and the second largest value among valuation profile $v$.
\end{definition}

In our analysis, we focus on the revenue performance in stage 2 of the model, as this is the stage in which reserves are set.
\section{A Differentially Private Approach}\label{sec:dp-approach}
In this section, we propose a simple differentially private (DP) approach to set a reserve price, which can be used to control the revenue and incentive compatibility in the two-stage model. All the proofs are deferred to Appendix~\ref{app:dp-missing-proofs}.

In the robust clearing price mechanism below, the seller collects historical bid profiles $\B=\{b^{(k)}\}_{k=1}^K$ from stage 1 and computes the clearing price $\hat{p}^c_\lambda$ to minimize the empirical clearing loss~\eqref{eq:empirical-clearing-price-multi-bidder}, given a parameter $\lambda$. In stage 2, for each auction, the seller adds a random noise $z$ i.i.d.\ sampled from a (CDF) distribution $F$ on the clearing price $\hat{p}^c_\lambda$, to set the reserve price. This mechanism, which we call the \emph{Differentially Private Robust Clearing Price (DP-RCP) Mechanism}, is summarized in Algorithm~\ref{alg:dp-clearing-price-multi-bidder}.

\begin{algorithm}
\SetKwInOut{Initialize}{Initialize}
\SetKwInOut{Output}{Output}
\Initialize{Sample bids (profile) $\{b^{(k)}\}_{k=1}^K$ from stage 1, and noise distribution $F$.}
\Output{Reserve price $r$ for each auction.}

Compute a clearing price $\hat{p}^c_\lambda$
\begin{eqnarray}\label{eq:empirical-clearing-price-multi-bidder}
\hat{p}^c_\lambda = \argmin_p \frac{1}{K}\sum_{k=1}^K \ell^c(p, b^{(k)}; \lambda)
\end{eqnarray}

Randomly draw a noise $z\sim F$.

$r \leftarrow \max(\hat{p}^c_\lambda + z, 0)$.
\caption{Differentially Private Robust Clearing Price Mechanism\label{alg:dp-clearing-price-multi-bidder}}
\end{algorithm}

Suppose each bid profile $b^{(k)}$ in stage 1 follows a distribution $B$.
Since we assume each stage contains a sufficiently large number of auctions, $\hat{p}_\lambda^c$ from~(\ref{eq:empirical-clearing-price-multi-bidder}) can be regarded as the optimal expected clearing price, associated with the distribution of bids $B$ and $\lambda$. 
Given this observation, the reserve price $r_\lambda(z, \beta)$ for $n$ bidders with $\beta=(\beta_1, \cdots, \beta_n)$ bidding strategies adopted in stage 1, given a noise $z$ and parameter $\lambda$, can be defined as:
\begin{eqnarray}\label{eq:expected-clearing-price-with-noise}
r_\lambda(z, \beta) = \max(p^c_\lambda(\beta) + z, 0).
\end{eqnarray}

\noindent
Similarly to \cite{Shen19}, we first characterize $p^c_\lambda(\beta)$ in the following.
\begin{proposition}\label{prop:clearing-price-multiple-bidders}
Suppose $0\leq \lambda \leq n$, for $n$ bidders such that each bidder $i$'s value is drawn from $D_i$ and uses strictly increasing $\beta_i$ bidding strategy, the optimal clearing price $p^c_\lambda(\beta)$ is the solution to the following equation \footnote{Since $\beta_i$ is weakly increasing, we denote $\beta_i^{-1}(p) := \inf\{v: \beta_i(v) \geq p\}$.},
\begin{eqnarray*}
\sum_{i=1}^n D_i(\beta_i^{-1}(p)) = n - \lambda
\end{eqnarray*}
\end{proposition}

We next characterize the expected revenue in stage 2, achieved by the DP-RCP mechanism. The following theorem shows that with high probability, the revenue loss caused by the noise in the DP-RCP mechanism will be bounded by $O(1/\eps)$, with high probability. The randomness in the result comes from the random noise $z$.

\begin{theorem}[Revenue Guarantee]\label{thm:revenue-multiple-bidders}
Given $0\leq \lambda \leq n$, noise $z\sim \mathtt{Lap}(0, 1/\eps)$, $D^{(1)}(\cdot)$ and $D^{(2)}(\cdot)$ are both $L$-Lipschitz. For any strictly increasing bidding strategies $\beta$ in the DP-RCP mechanism, we have
\begin{eqnarray}
\rev\left(r_\lambda(z, \beta)\right) \geq \rev(p^c_\lambda(\beta)) -\frac{(3L+4)\ln(1/\delta)}{\eps},
\end{eqnarray}
holds with probability at least $1-\delta$, where $p^c_\lambda(\beta)$ is the optimal expected clearing price defined in Eq.~(\ref{eq:optimal-expected-clearing-price})
\end{theorem}

We next characterize the IC-metric of each bidder $i$ in the following.

\begin{theorem}[IC-metric]\label{thm:ic-metric-multiple-bidders}
For $n$ bidders and any noise distribution $F$, the IC-metric of the DP-RCP mechanism satisfies,
\begin{eqnarray*}
\mathtt{DIC}_i = 1 -\eta\cdot\frac{\E_{v_i}\left[\int\limits_0^{v_i}G_i(m_i) f(m_i - p^*(1)) d\,m_i\right]}{\E_{v_i}[v_i\cdot G_i(v_i)]}
\end{eqnarray*}
where $\eta=\frac{\partial p^*(\alpha)}{\partial \alpha}\Big|_{\alpha=1}$ and $p^*(\alpha) = p^c_\lambda((\beta_i, \I_{-i}))$, where $\beta_i = \beta_i(v_i) = \alpha v_i, \forall v_i \in V$.
\end{theorem}

\noindent
Here $\eta$ captures the \emph{local sensitivity} of the clearing price to linear bid shading. We characterize $\eta$ in the following.
\begin{eqnarray}\label{eq:charac-local-sensitivity}
\eta=\frac{\partial p^*(\alpha)}{\partial \alpha}\Big|_{\alpha=1} = \frac{p^*(1) D'_i\left(p^*(1)\right)}{\sum_{j\neq i}D'_j(p^*(1))}	
\end{eqnarray}

\noindent
\textbf{Remark.} For the single bidder setting, $\eta = D^{-1}(1-\lambda)$, which is the $(1-\lambda)$th quantile of the value. For $n$ i.i.d bidders, i.e., $D_i = D$ for all $i\in[n]$, we have $\eta = p^*(1)/n = D^{-1}\left(1-\frac{\lambda}{n}\right)/n$. This naturally generalizes the single-bidder case. 
As $n$ grows large, the sensitivity $\eta$ goes to 0 as long as the bid distribution is bounded. This is intuitive, as a bidder's misreporting should have a negligible effect when there is a large number of competitors.

To implement the DP-RCP mechanism, we need to randomly draw a noise to add to the reserve for each auction, which may be undesirable in practice. To address this difficulty, we next introduce another approach which applies noise to historical bids used to fit the clearing price (one single time), rather than to the output of each auction. 

\section{A Smoothing Approach}\label{sec:smooth-approach}

In this section, we introduce a smoothing approach to set a \emph{deterministic} reserve price to make it robust to bidder misreports. In this method, we add a random noise to each bid and compute the clearing price based on the smoothed bids (in stage 1), as shown in Algorithm~\ref{alg:smooth-clearing-price}. We call this mechanism \emph{Smoothing Robust Clearing Price (sRCP) Mechanism}. %

\begin{algorithm}
\SetKwInOut{Initialize}{Initialize}
\SetKwInOut{Output}{Output}
\Initialize{Sample bids $\{b^{(k)}\}_{k=1}^K$, and noise distribution $F$.}
\Output{Reserve price $r$ for all auctions.}
\For{$k=1,\cdots, K$}
{Generate noise vector $z^{(k)} \sim F^n$}

Compute clearing price
\begin{eqnarray}\label{eq:smooth-clearing-price}
\hat{p}^c_\lambda = \argmin_p \frac{1}{K}\sum_{k=1}^K \ell^c(p, b^{(k)}+z^{(k)}; \lambda)
\end{eqnarray}

Set reserve price for all auctions, $r \leftarrow \hat{p}^c_\lambda$
\caption{Smoothing Robust Clearing Price Mechanism\label{alg:smooth-clearing-price}}
\end{algorithm}

Similarly to the DP approach, suppose each bid profile $b^{(k)}$ in stage 1 follows a distribution $B$, $\hat{p}_\lambda^c$ (in Eq.~\ref{eq:smooth-clearing-price}) can be represented as the optimal expected smoothed clearing price, associated with distribution of bids $B$, distribution of noise $F$ and $\lambda$. Therefore, the reserve price $r^s_\lambda(\beta)$ for $n$ bidders adopting $\beta=(\beta_1,\cdots,\beta_n)$ bidding strategies in stage 1, computed by this sRCP mechanism, can be defined as:
\begin{eqnarray}\label{eq:smooth-reserve-price}
r^s_\lambda(\beta) = \argmax_p \E_{v\sim D, z\sim F}\left[\ell^c(p, \beta(v) + z; \lambda)\right]
\end{eqnarray}

We can characterize $r^s_\lambda(\beta)$ in sRCP in the following.
\begin{proposition}\label{prop:opt-smooth-clearing-price}
Given $0 \leq \lambda \leq n$, the reserve price $r^s_\lambda(\beta)$ computed by sRCP mechanism for $n$ bidders is $0$ if $\sum_{i=1}^n\E_{v_i}[F(-\beta_i(v_i))] \geq n - \lambda$. Otherwise, it is the solution of price $p$ to
$\sum_{i=1}^n \int_{0}^1 F(p - \beta_i(v_i)) d D_i(v_i) = n -\lambda$.
\end{proposition}

The proof of the above Proposition is deferred to Appendix~\ref{app:smooth-approach}.
By monotonicity of $F$, it is straightforward to verify the uniqueness of the $r^s_\lambda(\beta)$.
Theorem~\ref{thm:smooth-revenue} characterizes the revenue guarantee for the sRCP mechanism.

\begin{theorem}[Revenue Guarantee]\label{thm:smooth-revenue}
Let function $\gamma(\cdot)$ be the \emph{inverse} of  function $\sum_{i=1}^n D_i(\beta_i^{-1}(\cdot))$. Suppose $0 \leq \lambda \leq n$, noise $z \sim \mathtt{Lap}(0, 1/\eps)$, $D^{(1)}(\cdot)$ and $D^{(2)}(\cdot)$ are $L$-Lipschitz, $\gamma$ is $\mu$-Lipschitz, and $\sum_{i=1}^n\E_{v_i}[F(-\beta_i(v_i))] \geq n- \lambda$. For any strictly increasing bidding strategies $\beta$ in the sRCP mechanism, we have
\begin{eqnarray*}
\rev(r^s_\lambda(\beta))\geq
\rev(p^c_\lambda(\beta)) - (6L+8) \cdot\frac{\sqrt{\mu \max(n-\lambda, \lambda)}}{\sqrt{\eps}}
\end{eqnarray*}
for any $\delta > 0$, where $p^c_\lambda(\beta)$ is the optimal (non-robust) clearing price defined in Eq.~(\ref{eq:optimal-expected-clearing-price}).
\end{theorem}

\begin{proof}
We first rewrite the expected revenue as, $\rev(r) = r - 2 r D^{(2)}(r) + r D^{(1)}(r) + \E[v^{(2)}]$. See Appendix~\ref{app:auxiliary-lemmas} for the detailed derivative of the above representation of the expected revenue. Then, by the Lipschitz assumption of $D^{(1)}$ and $D^{(2)}$, we can bound 
\begin{eqnarray}\label{eq:rev-reserve-bound}
\left|\rev(r^s_\lambda(\beta)) - \rev(p^c_\lambda(\beta))\right| \leq (4+3L)\cdot\left|p^c_\lambda(\beta) - r^s_\lambda(\beta)\right|.
\end{eqnarray}

Next, we will show how to bound $\left|r^s_\lambda(\beta) - p^c_\lambda(\beta)\right|$. %
To facilitate the proof, we provide another characterization of $r^s_\lambda(\beta)$ (see Lemma~\ref{lem:smooth-another-characterization-reserve} in Appendix~\ref{app:auxiliary-lemmas}), such that
$
\sum_{i=1}^n \int_{-\infty}^{\infty} D_i(\beta_i^{-1}(r^s_\lambda(\beta) - z)) f(z) dz = n - \lambda$.

We observe $\PP\left(|z| \leq \frac{\ln(1/\delta)}{\eps}\right) \geq 1- \delta$ for any $0 < \delta < 1$ when $z\sim \mathtt{Lap}(0, 1/\eps)$. Therefore, by the non-negativity and monotonicity of $D_i(\beta_i^{-1}(\cdot))$, we have
\begin{eqnarray*}
 (1-\delta)\sum_{i=1}^n D_i\left(\beta_i^{-1}\left(r^s_\lambda(\beta) - \frac{\ln(1/\delta)}{\eps}\right)\right) \leq \sum_{i=1}^n \int_{-\infty}^{\frac{\ln(1/\delta)}{\eps}} D_i(\beta_i^{-1}(r^s_\lambda(\beta) - z)) f(z) dz \leq  n - \lambda
\end{eqnarray*}

Thus, $r^s_\lambda(\beta) \leq \frac{\ln(1/\delta)}{\eps} + \gamma(\frac{n-\lambda}{1-\delta})$. On the other hand, since $D_i(\beta_i^{-1}(x)) \leq 1$ for any $x$, we have,
\begin{eqnarray*}
(1-\delta)\sum_{i=1}^n D_i\left(\beta_i^{-1}\left(r^s_\lambda(\beta) + \frac{\ln(1/\delta)}{\eps}\right)\right) + n\delta \geq \sum_{i=1}^n \int_{-\frac{\ln(1/\delta)}{\eps}}^{\infty} D_i(\beta_i^{-1}(r^s_\lambda(\beta) - z)) f(z) dz + n\delta\\
\geq  \sum_{i=1}^n \int_{-\infty}^{\infty} D_i(\beta_i^{-1}(r^s_\lambda(\beta) - z)) f(z) dz = n - \lambda.
\end{eqnarray*}
This implies
$r^s_\lambda(\beta) \geq \gamma(\frac{n-\lambda -n\delta}{1-\delta}) - \frac{\ln(1/\delta)}{\eps}$.
By the characterization of $p^c_\lambda(\beta)$ in Proposition~\ref{prop:clearing-price-multiple-bidders} and the definition of function $\gamma$, $p^c_\lambda(\beta)= \gamma(n-\lambda)$. Given the lower and upper bounds of $r^s_\lambda(\beta)$, we can bound,
\begin{eqnarray*}
\left|r^s_\lambda(\beta) - p^c_\lambda(\beta)\right| &\leq& \frac{\ln(1/\delta)}{\eps} + \max\left\{\left\vert\gamma(\frac{n-\lambda -n\delta}{1-\delta}) - \gamma(n-\lambda)\right\vert, \left\vert\gamma(\frac{n-\lambda}{1-\delta}) - \gamma(n-\lambda)\right\vert\right\}\\
&\leq&  \frac{\ln(1/\delta)}{\eps} + \frac{\mu\delta\max\{n-\lambda, \lambda\}}{1-\delta}.
\end{eqnarray*}

Setting $\delta = \frac{1}{1 + \sqrt{\mu\eps \max(n-\lambda, \lambda)}}$, we have $\frac{\ln(1/\delta)}{\eps} \leq \sqrt{\frac{\mu \max(n-\lambda, \lambda)}{\eps}}$ by the fact that $\ln(1+x) \leq x, \forall x\geq 0$. Combining Eq.~(\ref{eq:rev-reserve-bound}), we complete the proof.
\end{proof}

The above theorem shows that sRCP has better revenue guarantee compared with DP-RCP in terms of $\eps$ with additional assumption for $\gamma$ function, i.e., $O(1/\sqrt{\eps})$ revenue loss in sRCP v.s. $O(1/\eps)$ revenue loss in DP-RCP. This advantage is also verified through our simulations, and we also observe that it depends on the choice of $\lambda$. Astute readers may note that $\mu$ depends on number of bidders $n$, but it is decreasing with $n$ in general. Therefore, the revenue advantage achieved by sRCP mechanism is even stronger with a large number of bidders.
We characterize the IC-metric (see Definition~\ref{def:ic-metric}) in the following theorem, the proof is deferred to Appendix~\ref{app:smooth-approach}.

\begin{theorem}[IC-metric]\label{thm:smooth-ic-metric}
Let $\kappa = \sum_{i=1}^n \E_{v_i}[F(-v_i)]$.
For $n$ bidders and any noise distribution $F$, the IC-metric of the robust clearing price mechanism by smoothing approach satisfies,
\begin{eqnarray*}
\mathtt{DIC}_i = 1 - \frac{\zeta \cdot G_i(r^*(1)) \cdot (1- D_i(r^*(1)))}{\E_{v_i}[v_i \cdot G_i(v_i)]}\\
\mbox{ and } \zeta = \left\{
\begin{array}{cc}
\frac{\sum_{i=1}^n \int_0^1 v_i f(r^*(1) - v_i) d D_i(v_i)}{\sum_{i=1}^n \int_0^1 f(r^*(1) - v_i) d D_i(v_i)} & \mbox{ if } \kappa < n - \lambda\\
0 & \mbox{ if } \kappa > n - \lambda\\
\frac{\sum_{i=1}^n \int_0^1 v_i f(r^*(1) - v_i) d D_i(v_i)}{2\sum_{i=1}^n \int_0^1 f(r^*(1) - v_i) d D_i(v_i)} & \mbox{ if } \kappa = n - \lambda
\end{array}
\right.
\end{eqnarray*}
where $r^*(\alpha) = r^s_\lambda((\beta_i, \I_{-i}))$  when  $\beta_i(v_i) = \alpha v_i, \forall v_i\in V$ for any $\alpha > 0$.
\end{theorem}

\noindent\textbf{Remark.} $G_i(r^*(1)) \cdot (1- D_i(r^*(1)))$ is the probability that $r^*(1)$ is between $m_i$ and $v_i$. Indeed, $\zeta$ is similar to $\eta$ in Theorem~\ref{thm:ic-metric-multiple-bidders} and $\zeta = \frac{1}{2}\left[\frac{\partial r^*(\alpha)}{\partial \alpha}\Big|_{\alpha = 1^+} + \frac{\partial r^*(\alpha)}{\partial \alpha}\Big|_{\alpha = 1^-}\right]$. $\zeta$ depends on $\lambda$ \emph{implicitly}: when $\kappa > n -\lambda$, sRCP is dynamic-IC since $\mathtt{DIC}_i = 1$ for each bidder $i$; when $\kappa = n-\lambda$, $\zeta$ is just half of the $\zeta$ under $\kappa < n-\lambda$ and sRCP is more dynamic-IC. 

\section{Experiments}\label{sec:experiments}
In this section, we run simulations over synthetic data to validate our theoretical findings. By varying the magnitude $\eps$ of the noise and parameter $\lambda$ in the clearing loss, we obtain the tradeoff between revenue and IC-metric in the DP-RCP and sRCP mechanisms. %
\vspace{-5pt}
\subsection{Set-up}
For our simulations we assume the bids of each bidder follow a truncated log-normal distribution: $\mathtt{Lognormal}(0, 0.5)$ truncated by $[0, 2.5]$.\footnote{We truncate the bids to make them bounded to be consistent with our theory. Indeed, we observe the similar results without the truncation.} %
The lognormal is commonly used to model bid and value distributions in practice~\citep{thompson2013revenue}. We randomly generate 5,000 sample bids profiles for each stage.
To compute the IC-metric, we set the perturbation magnitude to $\alpha = 0.1$. The revenue is normalized by the expected total value in stage 2 assuming there is no reserve price and all bidders report truthfully. We only focus on Laplace noise distribution. All experiments are repeated 10 times and we present 95\% confidence intervals in all plots.

\subsection{Expected Revenue and IC-metric}
Due to space limitations, we only report the tradeoff between revenue and IC-metric for the single bidder setting and the two bidders setting.

\begin{figure}[h!]
\begin{subfigure}[b]{0.49\textwidth}
\centering
\includegraphics[scale=0.39]{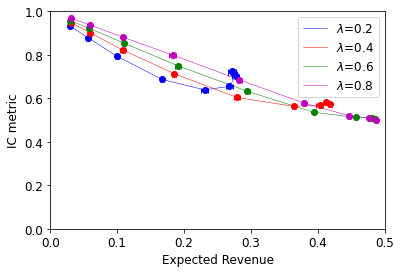}
\caption{The DP-RCP mechanism.}
\label{fig:dp-rev-ic-single-bidder-laplace}
\end{subfigure}
\begin{subfigure}[b]{0.49\textwidth}
\centering
\includegraphics[scale=0.39]{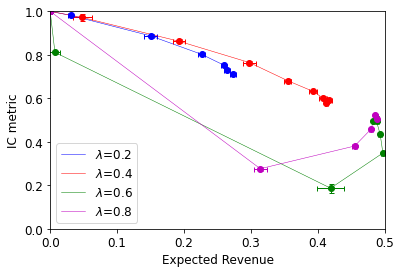}
\caption{The sRCP mechanism.}
\label{fig:smooth-rev-ic-single-bidder-laplace}
\end{subfigure}
\caption{Single bidder with $\mathtt{Lap}(0,1/\eps)$ noise distribution: tradeoff between revenue and IC-metric by varying $\lambda$ and $\eps$ for (a) the DP-RCP mechanism and (b) the sRCP mechanism.}
\label{fig:dp-single-bidder-laplace}
\end{figure}

For the single bidder setting, we exhibit the tradeoff between the expected revenue and IC-metric in Figure~\ref{fig:dp-rev-ic-single-bidder-laplace} for the \emph{DP-RCP} mechanism with different $\lambda =\{0.2, 0.4, 0.6, 0.8\}$ (as we observe that revenue is quite low when $\lambda = 0, 1$) and $\eps=\{0.1, 0.2, 0.4, 0.8, 1.6, 3.2, 6.4, 12.8, +\infty\}$ (no noise corresponds to $\eps=+\infty$). Based on this figure, the setting of $\lambda = 0.8$ (i.e., 20th quantile) in fact achieves the best tradeoff. %
When $\lambda$ is small, e.g. $\lambda = 0.2$, the clearing price is large and revenue achieved by the clearing price (with no noise) is small. In this case, adding a small noise (with large $\eps$) may decrease the IC-metric based on our characterization in Theorem~\ref{thm:ic-metric-multiple-bidders}, when $\eta$ is large. In addition, we test the tradeoff between the expected revenue and IC-metric for the \emph{sRCP} mechanism for the single bidder setting and we visualize it in Figure~\ref{fig:smooth-rev-ic-single-bidder-laplace}. We find that $\lambda=0.4$ (i.e., 60th quantile) achieves the best tradeoff between expected revenue and IC-metric, in the sense that at any desired level of expected revenue, $\lambda = 0.4$ can achieve the highest IC metric with an appropriately chosen noise level $\eps$. Indeed, sRCP achieves better tradeoff than DP-RCP when $\lambda=0.2, 0.4$, while DP-RCP performs better than sRCP when $\lambda$ is larger. Based on the experiments, sRCP mechanism prefers more aggressive setting (higher quantile to set clearing price, or equivalently lower $\lambda$) and DP-RCP mechanism benefits from more conservative (higher $\lambda$) settings.

\begin{figure}[h!]
\hspace{-15pt}
\begin{subfigure}[b]{0.49\textwidth}
\centering
\includegraphics[scale=0.4]{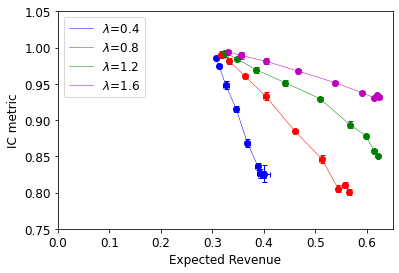}
\caption{The DP-RCP mechanism.}
\label{fig:dp-rev-ic-multiple-bidders-laplace}
\end{subfigure}
\begin{subfigure}[b]{0.49\textwidth}
\centering
\includegraphics[scale=0.4]{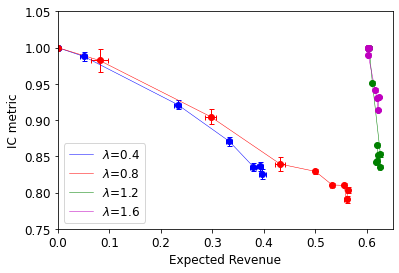}
\caption{The sRCP mechanism.}
\label{fig:smooth-rev-ic-multiple-bidders-laplace}
\end{subfigure}
\caption{Two bidders with $\mathtt{Lap}(0,1/\eps)$ noise distribution: tradeoff between revenue and IC-metric by varying $\lambda$ and $\eps$ for (a) the DP-RCP mechanism and (b) the sRCP mechanism.}
\label{fig:multiple-bidders-laplace}
\end{figure}

For the two bidders setting, we visualize the tradeoff between revenue and IC-metric of bidder 1 with different $\eps$s and $\lambda$s  for DP-RCP and sRCP in Figure~\ref{fig:dp-rev-ic-multiple-bidders-laplace} and Figure~\ref{fig:smooth-rev-ic-multiple-bidders-laplace}, respectively.\footnote{In this simulation, the two bidders are i.i.d. Therefore, we only report the IC-metric for bidder 1.}  As we observe, in the two bidders setting, with the existence of a competing bid, the IC-metric of bidder 1 is better than the IC-metric in the single bidder setting. The two RCP mechanisms both achieve better tradeoffs between revenue and IC-metric when $\lambda$ is sufficiently large. For the DP-RCP mechanism, Figure~\ref{fig:dp-rev-ic-multiple-bidders-laplace} shows that $\lambda=1.6$ achieves the best revenue-IC tradeoff among the set $\{0.4, 0.8, 1.2, 1.6\}$. Its tradeoff curve is remarkably better than the others, as it can achieve expected revenue close to the optimum with just an 0.05 decrease in the IC metric from 1.0 (the incentive compatible level). To reach this far, other settings see their IC metric dip below 0.9. For the sRCP mechanism, we observe $\lambda = 1.6$ achieves the best revenue-IC tradeoff, which is similar to the DP-RCP mechanism. Moreover, we find sRCP significantly outperforms than DP-RCP when $\lambda = 1.2, 1.6$.

\vspace{-10pt}
\section{Conclusion}
This work proposed two robust clearing price mechanisms (DP-RCP and sRCP) to set anonymous reserve prices based on historical bid profiles. The mechanisms are both simple to implement: the first approach fits a clearing price to bid data (which involves optimizing a convex loss function) and the price is made robust to misreports by agents by adding noise to the mechanism output, while the second approach computes a clearing price based on smoothed bid data (adding noise to the training bids). For both mechanisms, we provided bounds on the revenue guarantee in terms of the noise level applied.  %
We also provided exact characterizations of their IC-metric as defined in~\citep{DLMZ20}, which quantifies bidders' incentives to uniformly shade their bids. %
Our empirical evaluation showed that sRCP outperforms than DP-RCP in terms of the revenue and IC metric tradeoff with appropriately chosen parameter $\lambda$.

\bibliographystyle{ACM-Reference-Format}
\bibliography{robust-clearing-price}


\begin{thebibliography}{00}


\ifx \showCODEN    \undefined \def \showCODEN     #1{\unskip}     \fi
\ifx \showDOI      \undefined \def \showDOI       #1{{\tt DOI:}\penalty0{#1}\ }
  \fi
\ifx \showISBNx    \undefined \def \showISBNx     #1{\unskip}     \fi
\ifx \showISBNxiii \undefined \def \showISBNxiii  #1{\unskip}     \fi
\ifx \showISSN     \undefined \def \showISSN      #1{\unskip}     \fi
\ifx \showLCCN     \undefined \def \showLCCN      #1{\unskip}     \fi
\ifx \shownote     \undefined \def \shownote      #1{#1}          \fi
\ifx \showarticletitle \undefined \def \showarticletitle #1{#1}   \fi
\ifx \showURL      \undefined \def \showURL       #1{#1}          \fi
\providecommand\bibfield[2]{#2}
\providecommand\bibinfo[2]{#2}
\providecommand\natexlab[1]{#1}
\providecommand\showeprint[2][]{arXiv:#2}

\bibitem[\protect\citeauthoryear{Amin, Rostamizadeh, and Syed}{Amin
  et~al\mbox{.}}{2013}]%
        {amin2013learning}
\bibfield{author}{\bibinfo{person}{Kareem Amin}, \bibinfo{person}{Afshin
  Rostamizadeh}, {and} \bibinfo{person}{Umar Syed}.}
  \bibinfo{year}{2013}\natexlab{}.
\newblock \showarticletitle{Learning prices for repeated auctions with
  strategic buyers}. In \bibinfo{booktitle}{{\em Advances in Neural Information
  Processing Systems}}.
\newblock


\bibitem[\protect\citeauthoryear{Amin, Rostamizadeh, and Syed}{Amin
  et~al\mbox{.}}{2014}]%
        {amin2014repeated}
\bibfield{author}{\bibinfo{person}{Kareem Amin}, \bibinfo{person}{Afshin
  Rostamizadeh}, {and} \bibinfo{person}{Umar Syed}.}
  \bibinfo{year}{2014}\natexlab{}.
\newblock \showarticletitle{Repeated contextual auctions with strategic
  buyers}. In \bibinfo{booktitle}{{\em Advances in Neural Information
  Processing Systems}}.
\newblock


\bibitem[\protect\citeauthoryear{Balcan, Sandholm, and Vitercik}{Balcan
  et~al\mbox{.}}{2019}]%
        {balcan2019estimating}
\bibfield{author}{\bibinfo{person}{Maria-Florina Balcan},
  \bibinfo{person}{Tuomas Sandholm}, {and} \bibinfo{person}{Ellen Vitercik}.}
  \bibinfo{year}{2019}\natexlab{}.
\newblock \showarticletitle{Estimating Approximate Incentive Compatibility}. In
  \bibinfo{booktitle}{{\em Proceedings of the 2019 ACM Conference on Economics
  and Computation}} {\em (\bibinfo{series}{EC '19})}. \bibinfo{publisher}{ACM},
  \bibinfo{address}{New York, NY, USA}.
\newblock


\bibitem[\protect\citeauthoryear{Caillaud and Mezzetti}{Caillaud and
  Mezzetti}{2004}]%
        {Caillaud04}
\bibfield{author}{\bibinfo{person}{Bernard Caillaud} {and}
  \bibinfo{person}{Claudio Mezzetti}.} \bibinfo{year}{2004}\natexlab{}.
\newblock \showarticletitle{Equilibrium reserve prices in sequential ascending
  auctions}.
\newblock \bibinfo{journal}{{\em Journal of Economic Theory\/}}
  \bibinfo{volume}{117}, \bibinfo{number}{1} (\bibinfo{year}{2004}),
  \bibinfo{pages}{78 -- 95}.
\newblock


\bibitem[\protect\citeauthoryear{Cole and Roughgarden}{Cole and
  Roughgarden}{2014}]%
        {cole2014sample}
\bibfield{author}{\bibinfo{person}{Richard Cole} {and} \bibinfo{person}{Tim
  Roughgarden}.} \bibinfo{year}{2014}\natexlab{}.
\newblock \showarticletitle{The sample complexity of revenue maximization}. In
  \bibinfo{booktitle}{{\em Proceedings of the forty-sixth annual ACM symposium
  on Theory of computing}}.
\newblock


\bibitem[\protect\citeauthoryear{Colini-Baldeschi, Leonardi, Schrijvers, and
  Sodomka}{Colini-Baldeschi et~al\mbox{.}}{2020}]%
        {colini2020envy}
\bibfield{author}{\bibinfo{person}{Riccardo Colini-Baldeschi},
  \bibinfo{person}{Stefano Leonardi}, \bibinfo{person}{Okke Schrijvers}, {and}
  \bibinfo{person}{Eric Sodomka}.} \bibinfo{year}{2020}\natexlab{}.
\newblock \showarticletitle{Envy, regret, and social welfare loss}. In
  \bibinfo{booktitle}{{\em Proceedings of The Web Conference 2020}}.
\newblock


\bibitem[\protect\citeauthoryear{Deng, Lahaie, and Mirrokni}{Deng
  et~al\mbox{.}}{2019}]%
        {deng2019robustNon}
\bibfield{author}{\bibinfo{person}{Yuan Deng}, \bibinfo{person}{S{\'e}bastien
  Lahaie}, {and} \bibinfo{person}{Vahab Mirrokni}.}
  \bibinfo{year}{2019}\natexlab{}.
\newblock \showarticletitle{A Robust Non-Clairvoyant Dynamic Mechanism for
  Contextual Auctions}. In \bibinfo{booktitle}{{\em Advances in Neural
  Information Processing Systems}}.
\newblock


\bibitem[\protect\citeauthoryear{Deng, Lahaie, Mirrokni, and Zuo}{Deng
  et~al\mbox{.}}{2020}]%
        {DLMZ20}
\bibfield{author}{\bibinfo{person}{Yuan Deng}, \bibinfo{person}{S\'{e}bastien
  Lahaie}, \bibinfo{person}{Vahab Mirrokni}, {and} \bibinfo{person}{Song Zuo}.}
  \bibinfo{year}{2020}\natexlab{}.
\newblock \showarticletitle{A Data-Driven Metric of Incentive Compatibility}.
  In \bibinfo{booktitle}{{\em Proceedings of The Web Conference 2020}}.
\newblock


\bibitem[\protect\citeauthoryear{Deng, Lahaie, Mirrokni, and Zuo}{Deng
  et~al\mbox{.}}{2021}]%
        {DengEtAl21}
\bibfield{author}{\bibinfo{person}{Yuan Deng}, \bibinfo{person}{S\'{e}bastien
  Lahaie}, \bibinfo{person}{Vahab Mirrokni}, {and} \bibinfo{person}{Song Zuo}.}
  \bibinfo{year}{2021}\natexlab{}.
\newblock \showarticletitle{Revenue-Incentive Tradeoffs in Reserve Pricing}. In
  \bibinfo{booktitle}{{\em Proceedings of the 38th International Conference on
  Machine Learning (ICML-21), to appear}}.
\newblock


\bibitem[\protect\citeauthoryear{Devanur, Huang, and Psomas}{Devanur
  et~al\mbox{.}}{2016}]%
        {devanur2016sample}
\bibfield{author}{\bibinfo{person}{Nikhil~R Devanur}, \bibinfo{person}{Zhiyi
  Huang}, {and} \bibinfo{person}{Christos-Alexandros Psomas}.}
  \bibinfo{year}{2016}\natexlab{}.
\newblock \showarticletitle{The sample complexity of auctions with side
  information}. In \bibinfo{booktitle}{{\em Proceedings of the forty-eighth
  annual ACM symposium on Theory of Computing}}.
\newblock


\bibitem[\protect\citeauthoryear{Drutsa}{Drutsa}{2018}]%
        {drutsa2018weakly}
\bibfield{author}{\bibinfo{person}{Alexey Drutsa}.}
  \bibinfo{year}{2018}\natexlab{}.
\newblock \showarticletitle{Weakly Consistent Optimal Pricing Algorithms in
  Repeated Posted-Price Auctions with Strategic Buyer}. In
  \bibinfo{booktitle}{{\em International Conference on Machine Learning}}.
\newblock


\bibitem[\protect\citeauthoryear{Drutsa}{Drutsa}{2020}]%
        {drutsa2020reserve}
\bibfield{author}{\bibinfo{person}{Alexey Drutsa}.}
  \bibinfo{year}{2020}\natexlab{}.
\newblock \showarticletitle{Reserve pricing in repeated second-price auctions
  with strategic bidders}. In \bibinfo{booktitle}{{\em International Conference
  on Machine Learning}}.
\newblock


\bibitem[\protect\citeauthoryear{Duetting, Feng, Narasimhan, Parkes, and
  Ravindranath}{Duetting et~al\mbox{.}}{2019}]%
        {dutting19a}
\bibfield{author}{\bibinfo{person}{Paul Duetting}, \bibinfo{person}{Zhe Feng},
  \bibinfo{person}{Harikrishna Narasimhan}, \bibinfo{person}{David Parkes},
  {and} \bibinfo{person}{Sai~Srivatsa Ravindranath}.}
  \bibinfo{year}{2019}\natexlab{}.
\newblock \showarticletitle{Optimal Auctions through Deep Learning} {\em
  (\bibinfo{series}{Proceedings of Machine Learning Research})}.
\newblock


\bibitem[\protect\citeauthoryear{Epasto, Mahdian, Mirrokni, and Zuo}{Epasto
  et~al\mbox{.}}{2018}]%
        {epasto2018incentive}
\bibfield{author}{\bibinfo{person}{Alessandro Epasto},
  \bibinfo{person}{Mohammad Mahdian}, \bibinfo{person}{Vahab Mirrokni}, {and}
  \bibinfo{person}{Song Zuo}.} \bibinfo{year}{2018}\natexlab{}.
\newblock \showarticletitle{Incentive-aware learning for large markets}. In
  \bibinfo{booktitle}{{\em Proceedings of the 2018 World Wide Web Conference}}.
\newblock


\bibitem[\protect\citeauthoryear{Feng, Schrijvers, and Sodomka}{Feng
  et~al\mbox{.}}{2019}]%
        {feng2019online}
\bibfield{author}{\bibinfo{person}{Zhe Feng}, \bibinfo{person}{Okke
  Schrijvers}, {and} \bibinfo{person}{Eric Sodomka}.}
  \bibinfo{year}{2019}\natexlab{}.
\newblock \showarticletitle{Online Learning for Measuring Incentive
  Compatibility in Ad Auctions}. In \bibinfo{booktitle}{{\em The World Wide Web
  Conference}}.
\newblock


\bibitem[\protect\citeauthoryear{Golrezaei, Javanmard, and Mirrokni}{Golrezaei
  et~al\mbox{.}}{2019}]%
        {golrezaei2019dynamic}
\bibfield{author}{\bibinfo{person}{Negin Golrezaei}, \bibinfo{person}{Adel
  Javanmard}, {and} \bibinfo{person}{Vahab Mirrokni}.}
  \bibinfo{year}{2019}\natexlab{}.
\newblock \showarticletitle{Dynamic incentive-aware learning: Robust pricing in
  contextual auctions}. In \bibinfo{booktitle}{{\em Advances in Neural
  Information Processing Systems}}.
\newblock


\bibitem[\protect\citeauthoryear{Kanoria and Nazerzadeh}{Kanoria and
  Nazerzadeh}{2014}]%
        {kanoria2014dynamic}
\bibfield{author}{\bibinfo{person}{Yash Kanoria} {and} \bibinfo{person}{Hamid
  Nazerzadeh}.} \bibinfo{year}{2014}\natexlab{}.
\newblock \showarticletitle{Dynamic Reserve Prices for Repeated Auctions:
  Learning from Bids}. In \bibinfo{booktitle}{{\em Web and Internet Economics
  (WINE)}}.
\newblock


\bibitem[\protect\citeauthoryear{Liu, Huang, and Wang}{Liu
  et~al\mbox{.}}{2018}]%
        {liu2018learning}
\bibfield{author}{\bibinfo{person}{Jinyan Liu}, \bibinfo{person}{Zhiyi Huang},
  {and} \bibinfo{person}{Xiangning Wang}.} \bibinfo{year}{2018}\natexlab{}.
\newblock \showarticletitle{Learning optimal reserve price against non-myopic
  bidders}. In \bibinfo{booktitle}{{\em Advances in Neural Information
  Processing Systems}}.
\newblock


\bibitem[\protect\citeauthoryear{Maillé}{Maillé}{2007}]%
        {maille2007}
\bibfield{author}{\bibinfo{person}{Patrick Maillé}.}
  \bibinfo{year}{2007}\natexlab{}.
\newblock \showarticletitle{Market clearing price and equilibria of the
  progressive second price mechanism}.
\newblock \bibinfo{journal}{{\em RAIRO - Operations Research\/}}
  \bibinfo{volume}{41}, \bibinfo{number}{4} (\bibinfo{year}{2007}).
\newblock


\bibitem[\protect\citeauthoryear{Mao, Leme, and Schneider}{Mao
  et~al\mbox{.}}{2018}]%
        {maocontextual}
\bibfield{author}{\bibinfo{person}{Jieming Mao}, \bibinfo{person}{Renato~Paes
  Leme}, {and} \bibinfo{person}{Jon Schneider}.}
  \bibinfo{year}{2018}\natexlab{}.
\newblock \showarticletitle{Contextual Pricing for Lipschitz Buyers}.
\newblock  (\bibinfo{year}{2018}).
\newblock


\bibitem[\protect\citeauthoryear{McSherry and Talwar}{McSherry and
  Talwar}{2007}]%
        {mcsherry2007mechanism}
\bibfield{author}{\bibinfo{person}{Frank McSherry} {and} \bibinfo{person}{Kunal
  Talwar}.} \bibinfo{year}{2007}\natexlab{}.
\newblock \showarticletitle{Mechanism design via differential privacy}. In
  \bibinfo{booktitle}{{\em 48th Annual IEEE Symposium on Foundations of
  Computer Science (FOCS'07)}}. IEEE.
\newblock


\bibitem[\protect\citeauthoryear{Medina and Mohri}{Medina and Mohri}{2014}]%
        {medina2014learning}
\bibfield{author}{\bibinfo{person}{Andres~M Medina} {and}
  \bibinfo{person}{Mehryar Mohri}.} \bibinfo{year}{2014}\natexlab{}.
\newblock \showarticletitle{Learning theory and algorithms for revenue
  optimization in second price auctions with reserve}. In
  \bibinfo{booktitle}{{\em Proceedings of the 31st International Conference on
  Machine Learning (ICML-14)}}.
\newblock


\bibitem[\protect\citeauthoryear{Morgenstern and Roughgarden}{Morgenstern and
  Roughgarden}{2015}]%
        {morgenstern2015pseudo}
\bibfield{author}{\bibinfo{person}{Jamie~H Morgenstern} {and}
  \bibinfo{person}{Tim Roughgarden}.} \bibinfo{year}{2015}\natexlab{}.
\newblock \showarticletitle{On the pseudo-dimension of nearly optimal
  auctions}. In \bibinfo{booktitle}{{\em Advances in Neural Information
  Processing Systems}}.
\newblock


\bibitem[\protect\citeauthoryear{Myerson}{Myerson}{1981}]%
        {myerson1981optimal}
\bibfield{author}{\bibinfo{person}{Roger~B Myerson}.}
  \bibinfo{year}{1981}\natexlab{}.
\newblock \showarticletitle{Optimal auction design}.
\newblock \bibinfo{journal}{{\em Mathematics of Operations Research\/}}
  \bibinfo{volume}{6}, \bibinfo{number}{1} (\bibinfo{year}{1981}).
\newblock


\bibitem[\protect\citeauthoryear{Nedelec, El~Karoui, and Perchet}{Nedelec
  et~al\mbox{.}}{2019}]%
        {nedelec2019learning}
\bibfield{author}{\bibinfo{person}{Thomas Nedelec}, \bibinfo{person}{Noureddine
  El~Karoui}, {and} \bibinfo{person}{Vianney Perchet}.}
  \bibinfo{year}{2019}\natexlab{}.
\newblock \showarticletitle{Learning to bid in revenue-maximizing auctions}. In
  \bibinfo{booktitle}{{\em International Conference on Machine Learning}}.
\newblock


\bibitem[\protect\citeauthoryear{Ostrovsky and Schwarz}{Ostrovsky and
  Schwarz}{2011}]%
        {ostrovsky2011reserve}
\bibfield{author}{\bibinfo{person}{Michael Ostrovsky} {and}
  \bibinfo{person}{Michael Schwarz}.} \bibinfo{year}{2011}\natexlab{}.
\newblock \showarticletitle{Reserve prices in internet advertising auctions: A
  field experiment}. In \bibinfo{booktitle}{{\em Proceedings of the 12th ACM
  conference on Electronic commerce}}.
\newblock


\bibitem[\protect\citeauthoryear{Paes~Leme, Pal, and Vassilvitskii}{Paes~Leme
  et~al\mbox{.}}{2016}]%
        {paes2016field}
\bibfield{author}{\bibinfo{person}{Renato Paes~Leme}, \bibinfo{person}{Martin
  Pal}, {and} \bibinfo{person}{Sergei Vassilvitskii}.}
  \bibinfo{year}{2016}\natexlab{}.
\newblock \showarticletitle{A field guide to personalized reserve prices}. In
  \bibinfo{booktitle}{{\em Proceedings of the 25th international conference on
  world wide web}}.
\newblock


\bibitem[\protect\citeauthoryear{Parkes, Kalagnanam, and Eso}{Parkes
  et~al\mbox{.}}{2001}]%
        {parkes2001achieving}
\bibfield{author}{\bibinfo{person}{David~C. Parkes}, \bibinfo{person}{Jayant
  Kalagnanam}, {and} \bibinfo{person}{Marta Eso}.}
  \bibinfo{year}{2001}\natexlab{}.
\newblock \showarticletitle{Achieving Budget-balance with Vickrey-based Payment
  Schemes in Exchanges}. In \bibinfo{booktitle}{{\em Proceedings of the 17th
  International Joint Conference on Artificial Intelligence}} {\em
  (\bibinfo{series}{IJCAI'01})}. \bibinfo{publisher}{Morgan Kaufmann Publishers
  Inc.}, \bibinfo{address}{San Francisco, CA, USA}.
\newblock


\bibitem[\protect\citeauthoryear{Shen, Lahaie, and Paes~Leme}{Shen
  et~al\mbox{.}}{2019}]%
        {Shen19}
\bibfield{author}{\bibinfo{person}{Weiran Shen}, \bibinfo{person}{Sebastien
  Lahaie}, {and} \bibinfo{person}{Renato Paes~Leme}.}
  \bibinfo{year}{2019}\natexlab{}.
\newblock \showarticletitle{Learning to Clear the Market}. In
  \bibinfo{booktitle}{{\em Proceedings of the 36th International Conference on
  Machine Learning}}.
\newblock


\bibitem[\protect\citeauthoryear{Thompson and Leyton-Brown}{Thompson and
  Leyton-Brown}{2013}]%
        {thompson2013revenue}
\bibfield{author}{\bibinfo{person}{David~RM Thompson} {and}
  \bibinfo{person}{Kevin Leyton-Brown}.} \bibinfo{year}{2013}\natexlab{}.
\newblock \showarticletitle{Revenue optimization in the generalized
  second-price auction}. In \bibinfo{booktitle}{{\em Proceedings of the
  fourteenth ACM conference on Electronic commerce}}.
\newblock


\end{thebibliography}

\onecolumn
\clearpage
\appendix
\begin{center}
{
\Large
\textbf{
Robust Clearing Price Mechanisms for Reserve Price Optimization}
~\\
~\\	
Appendix
}
\end{center}
\section{Auxiliary Lemmas}\label{app:auxiliary-lemmas}

\subsection{Representation of the Expected Revenue}

\begin{lemma}\label{lem:expected-revenue}
The expected revenue in stage 2 can be represented as,
\begin{eqnarray*}
\rev(r) = r - 2 r D^{(2)}(r) + r D^{(1)}(r) + \E[v^{(2)}]
\end{eqnarray*}
\end{lemma}

\begin{proof}
\begin{eqnarray*}
\rev(r)
&=& \E_{v}\left[\max(v^{(2)}, r)\cdot \1\{v^{(1)} \geq r\}\right]\\
&=& \int_{r}^1 x d D^{(2)}(x) + r \PP(v^{(2)} \leq r < v^{(1)})\\
&=& 1 - r D^{(2)}(r) - \int_r^1 D^{(2)}dx + r(D^{(1)}(r) - B_2(r))\\
&=& r - r D^{(2)} + \int_r^1 (1- D^{(2)}(x))dx + r(D^{(1)}(r) - D^{(2)}(r))\\
&=& r - 2 r D^{(2)}(r) + r D^{(1)}(r) + \E[v^{(2)}]
\end{eqnarray*}

\end{proof}

\subsection{Another Characterization of Reserve Price in the sRCP Mechanism}

\begin{lemma}\label{lem:smooth-another-characterization-reserve}
Given $0 \leq \lambda \leq n$, the reserve price $r^s_\lambda(\beta)$ computed by sRCP mechanism for $n$ bidders is $0$ if $\sum_{i=1}^n\E_{v_i}[F(-\beta_i(v_i))] \geq n - \lambda$. Otherwise, it is the solution of price $p$ to
\begin{eqnarray*}
\sum_{i=1}^n \int_{-\infty}^{\infty} D_i(\beta_i^{-1}(p - z)) f(z) dz = n - \lambda,
\end{eqnarray*}
where $\beta_i^{-1}(x) = 0, \forall x \leq 0$.
\end{lemma}
\begin{proof}
\begin{eqnarray*}
&&\E_{v\sim D, z\sim F}[\ell^c(p, \beta(v) + z; \lambda)]\\
 &=& \sum_{i=1}^n \int_{-\infty}^\infty \int_{\beta_i^{-1}(p-z)}^1 (\beta_i(v) + z - p) d D_i(v) f(z)dz  + \lambda p\\
&=&  \sum_{i=1}^n \int_{-\infty}^\infty \int_{\beta_i^{-1}(p-z)}^1 \beta_i(v) d D_i(v) f(z)dz +  \sum_{i=1}^n \int_{-\infty}^\infty (z - p)\cdot\left(1 - D_i(\beta_i^{-1}(p - z))\right) f(z)dz + \lambda p
\end{eqnarray*}

Taking the gradient of above formula w.r.t. $p$, we have
\begin{eqnarray*}
\frac{\partial \E_{v, z}[\ell^c(p, \beta(v) +z; \lambda)]}{\partial p}
&=& -\sum_{i=1}^n\int_{-\infty}^\infty (1 - D_i(\beta_i^{-1}(p - z))) f(z)dz + \lambda\\
&=& \sum_{i=1}^n\int_{-\infty}^\infty D_i(\beta_i^{-1}(p - z)) f(z) dz - (n-\lambda)
\end{eqnarray*}

By integral by part, we have $\sum_{i=1}^n\int_{-\infty}^\infty D_i(\beta_i^{-1}(-z)) f(z) dz = \sum_{i=1}^n\E_{v_i}[F(-\beta_i(v_i))]$. Then if $\sum_{i=1}^n\E_{v_i}[F(-\beta_i(v_i))] \geq n -\lambda$, the reserve price is equal to $0$. Otherwise, setting the gradient to be zero, we complete the proof.

\end{proof}

\section{Missing Proofs from Section~\ref{sec:dp-approach}}\label{app:dp-missing-proofs}

\subsection{Proof of Proposition~\ref{prop:clearing-price-multiple-bidders}}

\begin{proof}
Taking the derivative of $\E_v\left[\ell^c(p, \beta(v); \lambda)\right]$ w.r.t $p$, we have
{\small
	\begin{align*}
	&\nabla_p\E_{v}[\ell^c(p, \beta(v); \lambda)] = \E_{v}\left[\nabla_p\ell^c(p, \beta(v); \lambda)\right]\\
	&= \E_v\left[\lambda - \sum_{i=1}^n\1\{\beta_i(v_i)\geq p\}\right] = \lambda - \E_v\left[\sum_{i=1}^n\1\{\beta_i(v_i)\geq p\}\right]\\
	&=\lambda - \sum_{i=1}^n (1 - D_i(\beta_i^{-1}(p))) = \sum_{i=1}^n D_i(\beta_i^{-1}(p)) - (n - \lambda)
	\end{align*}
}
Setting the gradient to be zero and by the first order condition, we complete the proof.
\end{proof}

\subsection{Proof of Theorem~\ref{thm:revenue-multiple-bidders}}
\begin{proof}
By Lemma~\ref{lem:expected-revenue}, the expected revenue in stage 2 is
\begin{eqnarray*}
\rev(r) = r - 2 r D^{(2)}(r) + r D^{(1)}(r) + \E[v^{(2)}]
\end{eqnarray*}

By the Lipschitzness of $D^{(1)}$ and $D^{(2)}$, we can bound the difference of the revenue under reserve prices $r_\lambda(z, \beta)$ and $p^c_\lambda(\beta)$, as follows,
\begin{eqnarray*}
\left|\rev(r_\lambda(z, \beta)) - \rev(p^c_\lambda(\beta)) \right| &\leq & (4+3L) |\max(p^c_\lambda(\beta)+z, 0) - p^c_\lambda(\beta)|
\\
&\leq& (4+3L)|z|,
\end{eqnarray*}

By the property of Laplace distribution, $\PP\left(|z|\leq \frac{\ln(1/\delta)}{\eps}\right) \geq 1-\delta$. Finally, we have
\begin{eqnarray*}
\PP\left(\left|\rev(r_\lambda(z, \beta)) - \rev(p^c_\lambda(\beta)) \right|\leq (4+3L)\cdot \frac{\ln(1/\delta)}{\eps}\right)\geq \PP\left(|z|\leq \frac{\ln(1/\delta)}{\eps}\right) \geq 1-\delta
\end{eqnarray*}

\end{proof}

\subsection{Proof of Theorem~\ref{thm:ic-metric-multiple-bidders}}
\begin{proof}
We first rewrite the expected utility function $\E_{v_i}[\hat{u}_{i, 2}(\alpha; v)]$ in the following way,
\begin{eqnarray*}
&&\E_{v}[\hat{u}_{i, 2}(\alpha; v)]\\
&=& \E_v\left[\E_z\left[\1\{v_i \geq \max_{j\neq i}\{v_j, p^*(\alpha) + z\}\}\cdot (v_i - \max_{j\neq i}\{v_j, p^*(\alpha) + z\})\right]\right]\\
&=& \E_{v: v_i \geq m_i}\left[\int_{-\infty}^{m_i - p^*(\alpha)}(v_i - m_i) f(z) dz + \int_{m_i - p^*(\alpha)}^{v_i - p^*(\alpha)} (v_i - p^*(\alpha) - z) f(z) dz\right]\\
&=&\E_{v: v_i \geq m_i}\left[(v_i - m_i) F(m_i - p^*(\alpha)) + (v_i - p^*(\alpha))(F(v_i - p^*(\alpha)) - F(m_i - p^*(\alpha))) - \int_{m_i - p^*(\alpha)}^{v_i - p^*(\alpha)} z f(z) dz\right]\\
&=& \E_{v: v_i \geq m_i} \left[(v_i - p^*(\alpha))F(v_i - p^*(\alpha)) - (m_i - p^*(\alpha))F(m_i - p^*(\alpha)) - \int_{m_i - p^*(\alpha)}^{v_i - p^*(\alpha)} z f(z) dz\right]
\end{eqnarray*}

For notation simplicity, we denote $$T(v_i, m_i, \alpha) = \left[(v_i - p^*(\alpha))F(v_i - p^*(\alpha)) - (m_i - p^*(\alpha))F(m_i - p^*(\alpha)) - \int\limits_{m_i - p^*(\alpha)}^{v_i - p^*(\alpha)} z f(z) dz\right]\cdot g_i(m_i)$$

By Leibniz integral rule, we compute the quantity $\lim_{\alpha\rightarrow 0} \frac{\E_v[\hat{u}_{i,2}(1+\alpha; v)] - \E_v[\hat{u}_{i,2}(1-\alpha; v)]}{2\alpha}$ as follows,

\begin{eqnarray*}
&&\lim_{\alpha\rightarrow 0} \frac{\E_v[\hat{u}_{i,2}(1+\alpha; v)] - \E_v[\hat{u}_{i,2}(1-\alpha; v)]}{2\alpha}\\
&=& \frac{\partial \E_{v_i}\left[\int\limits_{0}^{v_i}T(v_i, m_i, \alpha)d\,m_i\right]}{\partial \alpha}\Bigg|_{\alpha=1}\\
&=& \E_{v_i}\left[v_i\cdot T(v_i, v_i, \alpha) + \int\limits_{0}^{v_i} \frac{\partial T(v_i, m_i, \alpha)}{\partial \alpha} d\,m_i\Bigg|_{\alpha=1}\right]\\
&=& \E_{v_i}\left[0 + \int\limits_{0}^{v_i}\left(F(v_i - p^*(\alpha))\cdot\frac{\partial [v_i - p^*(\alpha)]}{\partial \alpha} - F(m_i - p^*(\alpha))\cdot \frac{\partial [m_i - p^*(\alpha)]}{\partial \alpha}\right) g_i(m_i) d\,m_i\Bigg|_{\alpha=1}\right]\\
&& (\text{By the fact that } T(v_i, v_i, \alpha) = 0)\\
&=& \E_{v_i}\left[0 + \int\limits_{0}^{\alpha v_i}\left(F(v_i - p^*(\alpha))\cdot \left(-\frac{\partial p^*(\alpha)}{\partial \alpha}\right) + F(m_i - p^*(\alpha))\cdot \frac{\partial p^*(\alpha)}{\partial \alpha}\right) g_i(m_i) d\,m_i\Bigg|_{\alpha=1}\right]\\
&=& \E_{v_i}\left[\int\limits_{0}^{v_i}\left[\left(F(m_i - p^*(1)) - F(v_i - p^*(1))\right) \cdot \eta\right] g_i(m_i) d\,m_i\right]
\end{eqnarray*}

Therefore, we have
\begin{eqnarray*}
\mathtt{IC}_i &=& \lim_{\alpha\rightarrow 0} \frac{\E_v[\hat{u}_{i,2}(1+\alpha; v)] - \E_v[\hat{u}_{i,2}(1-\alpha; v)]}{2\alpha\E_{v_i}[v_i x_i(v_i)]}\\
&=& -\frac{\eta\cdot\E_{v_i}\left[\int\limits_0^{v_i} (F(v_i - p^*(1)) - F(m_i - p^*(1))) g_i(m_i)d\,m_i\right]}{\E_{v}[v_i x_{i, 1}(v)]}\\
&=& -\eta\cdot\frac{\E_{v_i}\left[ (F(v_i - p^*(1))G_i(v_i) - \int\limits_0^{v_i}F(m_i - p^*(1))) d\,G_i(m_i)\right]}{\E_{v}[v_i x_{i, 1}(v)]}\\
&=& -\eta\cdot\frac{\E_{v_i}\left[F(v_i - p^*(1))G_i(v_i) -F(m_i - p^*(1))G_i(m_i)\Big|_{0}^{v_i} + \int\limits_0^{v_i}G_i(m_i) f(m_i - p^*(1)) d\,m_i\right]}{\E_{v}[v_i x_{i, 1}(v)]}\\
&&(\text{By intergral by part})\\
&=& -\eta\cdot\frac{\E_{v_i}\left[\int\limits_0^{v_i}G_i(m_i) f(m_i - p^*(1)) d\,m_i\right]}{\E_{v_i}[v_i\cdot G_i(v_i)]}\\
\end{eqnarray*}
where the last equality above holds because
\begin{eqnarray*}
x_{i, 1}(v) = \PP(m_i \leq v_i) = G_i(v_i).
\end{eqnarray*}
\end{proof}

\subsection{Proof of Equation~(\ref{eq:charac-local-sensitivity})}
\begin{proof}
Given the definition of $p^*(\alpha)$ and Proposition~\ref{prop:clearing-price-multiple-bidders}, we have
$D_i\left(\frac{p^*(\alpha)}{\alpha}\right) + \sum_{j\neq i} D_j(p^*(\alpha)) = n - \lambda
$. Taking gradient w.r.t $\alpha$ of the both sides in the above equation, we have
{\small
\begin{eqnarray*}
\frac{\partial p^*(\alpha)}{\partial \alpha} = \frac{p^*(\alpha) D'_i\left(\frac{p^*(\alpha)}{\alpha}\right)}{\alpha D'_i\left(\frac{p^*(\alpha)}{\alpha}\right) + \alpha^2 \sum_{j\neq i}D'_j(p^*(\alpha))}.
\end{eqnarray*}
}
Therefore, $\eta = \frac{p^*(1) D'_i\left(p^*(1)\right)}{D'_i\left(p^*(1)\right) + \sum_{j\neq i}D'_j(p^*(1))}$
\end{proof}

\section{Missing Proofs from Section~\ref{sec:smooth-approach}}\label{app:smooth-approach}

\subsection{Proof of Proposition~\ref{prop:opt-smooth-clearing-price}}

\begin{proof}
\begin{eqnarray*}
\E_{v\sim D, z\sim F}[\ell^c(p, \beta(v) + z; \lambda)]
&=& \sum_{i=1}^n\int_{0}^1 \int_{p - \beta_i(v)}^\infty (\beta_i(v) + z - p) d F(z) d D_i(v)  + \lambda p\\
&=& \sum_{i=1}^n \int_0^1 (\beta_i(v) - p) \cdot (1 - F(p - \beta_i(v))) d D_i(v) \\
&&+ \sum_{i=1}^n \int_0^1 \int_{p-\beta_i(v)}^\infty z d F(z) d D_i(v) + \lambda p
\end{eqnarray*}

Taking the partial gradient of the above formula w.r.t $p$, we have
\begin{eqnarray*}
\frac{\partial \E_{v, z}[\ell^c(p, \beta(v) +z; \lambda)]}{\partial p}
&=& -\sum_{i=1}^n\int_0^1 (1-F(p-\beta_i(v))) dD_i(v) + \lambda\\
&=& \sum_{i=1}^n\int_0^1 F(p-\beta_i(v)) dD_i(v) - (n-\lambda)
\end{eqnarray*}

If $\sum_{i=1}^n\E_{v_i}[F(-\beta_i(v_i))] \geq n -\lambda$, the reserve price is equal to $0$. Otherwise, setting the gradient to be zero, we complete the proof. %
\end{proof}

%

%
%
%
%
%
%

%
%
%
%

%
%
%
%

%
%
%
%

%
%
%
%

%
%
%
%

%
%
%
%
%

%
%
%
%

%
%
%
%
%
%

\subsection{Proof of Theorem~\ref{thm:smooth-ic-metric}}
\begin{proof}
\begin{eqnarray*}
\E_{v_i}[\hat{u}_{i, 2}(\alpha; v)]
&=& \E_v\left[\1\{v_i \geq \max_{j\neq i}\{v_j, r^*(\alpha)\}\}\cdot (v_i - \max_{j\neq i}\{v_j, r^*(\alpha)\})\right]\\
&=& \E_{v: v_i \geq m_i, v_i \geq r^*(\alpha)}\left[(v_i - m_i)\cdot \1\{m_i \geq r^*(\alpha)\} + (v_i - r^*(\alpha))\cdot\1\{m_i \leq r^*(\alpha) \leq v_i\}\right]\\
&=& \E_{v_i: v_i \geq r^*(\alpha)}\left[\int_{r^*(\alpha)}^{v_i}(v_i - m_i)g_i(m_i) d m_i\right] + \int_{r^*(\alpha)}\int_0^{r^*(\alpha)} (v_i - r^*(\alpha)) g_i(m_i) dm_i d D_i(v_i)\\
&=& \E_{v_i: v_i \geq r^*(\alpha)}\left[\int_{r^*(\alpha)}^{v_i}(v_i - m_i)g_i(m_i) d m_i\right] + \int_{r^*(\alpha)} (v_i - r^*(\alpha)) G_i(r^*(\alpha)) dD_i(v_i)\\
\end{eqnarray*}

Therefore, we have
\begin{eqnarray*}
&&\lim_{\alpha\rightarrow 0} \frac{\E_v[\hat{u}_{i,2}(1+\alpha; v)] - \E_v[\hat{u}_{i,2}(1-\alpha; v)]}{2\alpha}\\
&=& \frac{1}{2}\cdot \left(\frac{\partial \E_v[\hat{u}_{i,2}(\alpha; v)]}{\partial \alpha}\Big|_{\alpha = 1^+} + \frac{\partial \E_v[\hat{u}_{i,2}(\alpha; v)]}{\partial \alpha}\Big|_{\alpha = 1^-}\right)\\
&=& -\E_{v_i: v_i \geq r^*(1)}\left[(v_i - r^*(1)) g_i(r^*(1))\cdot \zeta\right] + \E_{v_i: v_i \geq r^*(1)}\left[-\eta G_i(r^*(1)) + (v_i - r^*(1))g_i(r^*(1))\zeta\right]\\
&=& -\zeta\E_{v_i: v_i \geq r^*(1)}\left[G_i(r^*(1))\right],
\end{eqnarray*}
where $\zeta = \frac{1}{2}\left[\frac{\partial r^*(\alpha)}{\partial \alpha}\Big|_{\alpha = 1^+} + \frac{\partial r^*(\alpha)}{\partial \alpha}\Big|_{\alpha = 1^-}\right]$.
Therefore, the IC-metric for bidder $i$ is 
\begin{eqnarray*}
\mathtt{DIC}_i = 1 - \frac{\zeta\cdot G_i(r^*(1))\cdot (1-D_i(r^*(1)))}{\E_{v_i}[v_i \cdot G_i(v_i)]}
\end{eqnarray*}

Then we can derive $\eta$ in the following way, by Proposition~\ref{prop:opt-smooth-clearing-price}, when $\sum_{i=1}^n \E_{v_i}[F(-v_i)] < n - \lambda$, we have
\begin{eqnarray*}
\sum_{i=1}^n \int_{0}^1 F(r^*(\alpha) - \alpha v_i) d D_i(v_i) = n -\lambda
\end{eqnarray*}
Taking derivative with respect to $\alpha$ in the both sides, we have
\begin{eqnarray*}
\sum_{i=1}^n \int_0^1 f(r^*(\alpha) - \alpha v_i) \cdot \left(\frac{\partial r^*(\alpha)}{\partial \alpha} - v_i\right) d D_i(v_i) = 0
\end{eqnarray*}
Thus, we get
\begin{eqnarray*}
\frac{\partial r^*(\alpha)}{\partial \alpha}\Big|_{\alpha=1} = \frac{\sum_{i=1}^n \int_0^1 v_i f(r^*(1) - v_i) d D_i(v_i)}{\sum_{i=1}^n \int_0^1 f(r^*(1) - v_i) d D_i(v_i)}
\end{eqnarray*}

When $\sum_{i=1}^n \E_{v_i}[F(-v_i)] > n - \lambda$, there exists a $\delta > 0$, $r^*(\alpha) = 0, \forall \alpha \in [1- \delta, 1 + \delta]$. Thus $\zeta = 0$.

When $\sum_{i=1}^n \E_{v_i}[F(-v_i)] = n - \lambda$, the left derivative of $r^*(\alpha)$ at $\alpha = 1$ is 0, the right derivative of $r^*(\alpha)$ at $\alpha = 1$ is $\frac{\sum_{i=1}^n \int_0^1 v_i f(r^*(1) - v_i) d D_i(v_i)}{\sum_{i=1}^n \int_0^1 f(r^*(1) - v_i) d D_i(v_i)}$. Then $\zeta = \frac{\sum_{i=1}^n \int_0^1 v_i f(r^*(1) - v_i) d D_i(v_i)}{2\sum_{i=1}^n \int_0^1 f(r^*(1) - v_i) d D_i(v_i)}$.

Therefore, we complete the proof.

\end{proof}

\end{document}